\newcommand{\CLASSINPUTinnersidemargin}{20mm}
\documentclass[a4paper, conference]{IEEEtran}

\usepackage{graphicx}
\graphicspath{{./figs/}}

\usepackage{algorithm}
\usepackage{algorithmic}
\usepackage{amsmath}
\usepackage{amsfonts}
\usepackage{subfigure}
\usepackage{booktabs}
\usepackage{multirow}
\usepackage{color}

\begin{document}

\title{
Lithography Hotspot Detection and Mitigation in Nanometer VLSI}

\author{
\authorblockN{Jhih-Rong Gao, Bei Yu, Duo Ding, and David Z. Pan\\}
\authorblockA{Dept. of ECE, The University of Texas at Austin, Austin, TX 78712\\}
Email: dpan@mail.utexas.edu
}

\maketitle

\newtheorem{problem}{\textbf{Problem}}
\newtheorem{define}{\textbf{Definition}}
\newtheorem{theorem}{\textbf{Theorem}}
\newtheorem{lemma}{\textbf{Lemma}}
\newtheorem{conjecture}{Conjecture}

\begin{abstract}
With continued feature size scaling, even state of the art semiconductor manufacturing processes will often run into layouts with poor printability and yield. Identifying lithography hotspots is important at both physical verification and early physical design stages. While detailed lithography simulations can be very accurate, they may be too computationally expensive for full-chip scale and physical design inner loops. Meanwhile, pattern matching and machine learning based hotspot detection methods can provide acceptable quality and yet fast turn-around-time for full-chip scale physical verification and design. In this paper, we discuss some key issues and recent results on lithography hotspot detection and mitigation in nanometer VLSI.

\end{abstract}

\section{Introduction}
The continued shrinking of feature size has made IC manufacturing more and more prone to lithography hotspots, i.e., the layouts with poor printability and yield. To address this challenge, various resolution enhancement techniques (RETs), such as optical proximity correction (OPC), source mask co-optimization, and so on, have been proposed to improve pattern printability. However, lithography hotspots still remain a key challenging issue for IC designs in deep sub-wavelength processes (e.g., below 32nm), as the current lithography wavelength is still stuck at 193nm, which is much bigger than the feature size (e.g., 22nm). 

In conventional design flow, lithography simulations \cite{kim_spie03, roseboom_spie07}  have been used to identify problematic patterns. Lithography simulation is accurate but extremely computational intensive, especially for full-chip scale. If the lithography hotspot detection needs to be fed into some physical design stage to guide lithography-friendly layout optimization, it would be almost impossible to apply these detailed lithography simulations in the inner loop.

Recently, several hotspot detection approaches have been proposed, mainly based on pattern matching and machine learning techniques to avoid CPU-intensive lithography simulations. The challenges are how to extract critical information of these hotspot patterns and match them in the full-chip scale with high fidelity and low false alarm. There are also studies on integrating hotspot detection into physical design.  This paper will discuss some key aspects of these lithography hotspot detection and mitigation methods§.
 
The rest of the paper will be organized as follows. In Section \ref{sec:detection}, we discuss several lithography hotspot detection techniques. We then discuss hotspot mitigation in Section \ref{sec:mitigation}, followed by the conclusion in Section \ref{sec:conclude}.

\section{Lithography Hotspot Detection} \label{sec:detection}
\subsection{Layout Encoding Techniques} \label{sec:encode}
A hotspot is caused not only by a particular pattern, but also by the interaction with neighboring patterns inside the lithography influence region. One fundamental step for the hotspot detection in both pattern matching and machine learning methods is to represent layout patterns with certain format that can well describe the layout environment. Several layout encoding methods have been proposed to extract critical layout information from different aspects.

\begin{figure}[hb]
 \centering
\includegraphics[width=3.4in]{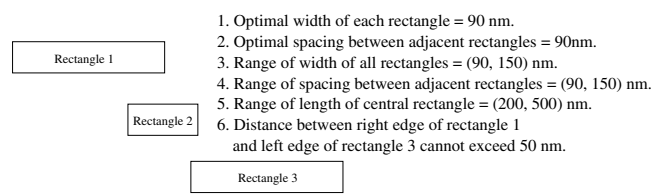}
\vspace{-0.1in}
  \caption{An example of range pattern staircase \cite{HOT_ICCAD07_Xu}.}
  \label{figure:range_pattern}
\end{figure}

The concept of range pattern \cite{HOT_ICCAD06_Yao} is proposed to incorporate process-dependent specifications, and is enhanced in \cite{HOT_ICCAD07_Xu} to represent new types of hotspots. A range pattern is a two-dimensional layout of rectangles with additional specifications encoded by strings. Fig. \ref{figure:range_pattern} shows an example of range pattern ``Staircase." Each range pattern is associated with a scoring mechanism to reflect the problematic regions according to yield impact. The hotspot patterns are stored in a pre-defined library and the detection process performs string matching to find hotspots. This approach is accurate, but the construction of range  patterns relies on a grid-based layout matrix, and may be time-consuming when the number of grids is large.

\begin{figure}[htbp]
 \centering
   \subfigure[]{\includegraphics[width=1in]{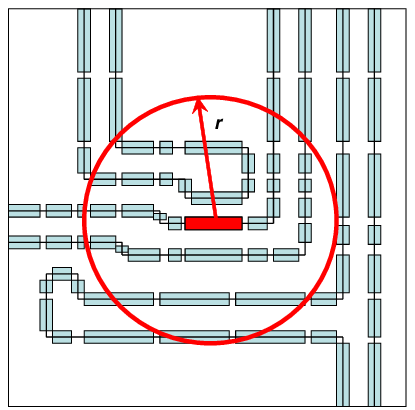}}
  \hspace{.16in}
  \subfigure[]{\includegraphics[width=1in]{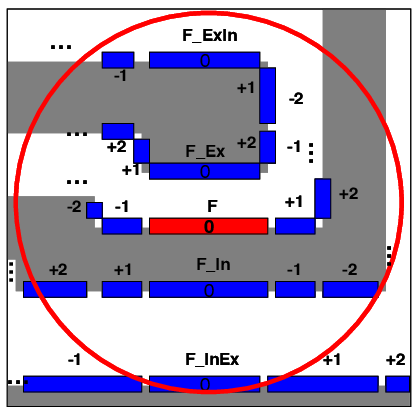}}
\vspace{-0.1in}
  \caption{Fragmentation based hotspot signature extraction \cite{HOT_ASPDAC2011_Ding}.}
  \label{figure:fragment}
\end{figure}

The context characterization \cite{HOT_ASPDAC2011_Ding} cuts a layout pattern into fragments. For each fragment $F$, an effective radius $r$ is defined to cover the neighboring fragments which need to be considered in the context characterization of $F$ as shown in Fig. \ref{figure:fragment}. A complete representation of $F$ includes the geometric characteristic of fragments inside $r$, including pattern shapes, the distance between patterns, corner information (convex or concave), and so on.

\begin{figure}[htbp]
 \centering
\includegraphics[width=3.2in]{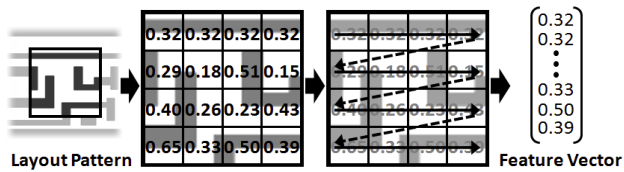}
\vspace{-0.1in}
  \caption{Density-based pattern representation. \cite{HOT_dac13_lin}.}
  \label{figure:density}
\end{figure}

The density-based pattern encoding is introduced in \cite{HOT_SPIE09_Wuu}, where a layout pattern is represented as a vector of layout density values of its surrounding area. Given a layout clip with predefined grids, the method calculates the layout density covered in each grid. An ordered list of density values then forms the final vector that represents the corresponding layout pattern. Fig. \ref{figure:density} illustrates the process of pattern encoding.
The goal of this representation is not to identify the geometrical features that may degrade the printability of a pattern. Instead, it aims at providing a compact representation of layout patterns to enable measurement of pattern similarities.

\subsection{Pattern Matching Based Hotspot Detection} \label{sec:pm}
In pattern matching based approaches, a set of known hotspots are pre-characterized and stored in a database. The hotspot detection process involves matching the tested layout patterns with the hotspots in the database. This method is fast and accurate at detecting known patterns, but it lacks the capability of predicting unseen data.

A layout graph is proposed in \cite{HOT_SPIE06_Kahng} to reflect pattern-related CD variation. The resulted graph can be used to find hotspots including closed features, L-shaped features and complex patterns. Yu et al. proposed a DRC-based hotspot detection \cite{HOT_DAC2012_Yu} by extracting critical topological features and modeling them as design rules. Therefore, hotspot detection can be viewed as a rule checking process through a DRC engine.

\begin{figure}[tbp]
  \centering
  \includegraphics[width=0.5\textwidth]{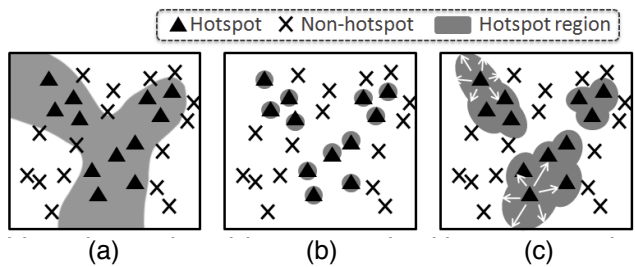}
  \caption{A 2D-space example of hotspot region decision. (a) Machine learning; (b) Pattern matching; (c) Fuzzy matching model. \cite{HOT_dac13_lin}}
  \label{fig:fuzzyMatch}
  \vspace{-.1in}
\end{figure}

Recently a fuzzy matching model was proposed in \cite{HOT_dac13_lin} which can dynamically tune appropriate fuzzy regions around known hotspots in multi-dimensional space. 
Fig. \ref{fig:fuzzyMatch} shows an example with known layout patterns of hotspots and non-hotspots in a 2-dimensional space. A machine learning method would divide the space into two regions of hotspots and non-hotspots as shown in Fig. \ref{fig:fuzzyMatch}(a), while a conventional pattern matching approach would construct an individual pattern to match each known hotspot as shown in (b). The fuzzy matching model in Fig. \ref{fig:fuzzyMatch}(c) includes groups of hotspots, where the fuzzy region of each group will iteratively grows to provide better detection accuracy.

\subsection{Machine Learning Based Hotspot Detection} \label{sec:ml}
Machine learning techniques construct a regression model based on a set of training data. This method can naturally identify previous unknown hotspots. However, it may generate false alarms, which are not real hotspots. How to improve the detecting accuracy is the main challenge when adopting machine learning techniques.

Many recent approaches utilize support vector machine (SVM) and artificial neural network (ANN) techniques to construct the hotspot detection kernel.
In \cite{HOT_DAC09_Drmanac}, a 2-D distance transform and histogram extraction is performed on pixel-based layout images, which are then used to construct the SVM-based hotspot detection.
In \cite{HOT_ASPDAC2011_Wuu}\cite{HOT_dac13_yu}, SVM is employed through extraction and classification of layout density-based metrics.
A neural network judgment based detection flow is proposed in \cite{HOT_ICICDT09_Ding}, where 2-D hotspot patterns are directly used to train an ANN kernel. A hybrid method \cite{HOT_TCAD2011_Ding} that adopts both SVM and ANN is presented to further improve the performance.

\subsection{Hybrid Machine Learning and Patterning Matching} \label{sec:hybrid}

\begin{figure}[tbp]
  \centering
  \includegraphics[width=3in]{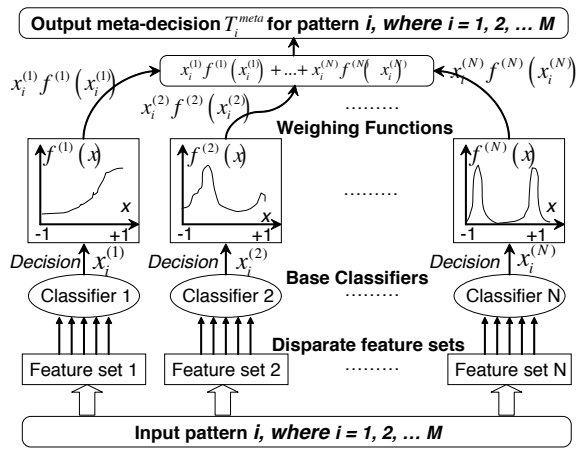}
  \caption{Meta-classifier construction via a combination of disparate base classifiers \cite{HOT_ASPDAC2012_Ding}.}
  \label{fig:meta_classifier}
  \vspace{-.1in}
\end{figure}

Since both patterning matching and machine learning have pros and cons, it will be  is to apply both machine learning models and pattern matching models. 
In \cite{HOT_ASPDAC2012_Ding}, data samples are fed to a pattern matcher first, then machine learning classifiers are used to examine the non-hotspots left by the pattern matcher. Motivated by the fact that different hotspot classifiers have different objectives and strengths, \cite{HOT_ASPDAC2012_Ding} further proposed a unified meta-classifier that enables several classifiers to work together. The meta-classifier is composed of multiple \textit{base classifiers} and \textit{weighting functions}. Each base classifier is an individual hotspot classifier that is optimized under certain performance metric, such as detection accuracy, false-alarms, adaptivity to new unknown designs, etc. Weighting functions are used to control the overall combination of base classifiers, which needs to be optimized to achieve better accuracy and less noise. 
The construction flow of meta-classification is illustrated in Fig. \ref{fig:meta_classifier}. For each layout pattern, certain hotspot features are extracted and then fed into each base classifier, which calculates the prediction decision and generates a weight based on the weighting functions. The final meta-decision is based on the weighed sum of base classifiers.

\subsection{Clustering in Hotspot Detection} \label{sec:cluster}
Since there are many design rules to guide and restrict physical design, in general the number of non-hotpot patterns greatly outnumber that of real hotspot patterns \cite{HOT_ICCAD2012_Torres}.
The imbalance between hotspot and non-hotspot data is called \textit{imbalanced populations}, and it may cause performance degradation for some hotspot detection approaches. In addition, how to avoid redundant data and reduce the time for detection is an important issue.

In \cite{HOT_SPIE08_Ma}\cite{HOT_SPIE09_Ghan}, the extracted hotspots are classified into clusters by data mining methods. An incremental clustering algorithm \cite{HOT_SPIE09_Ghan} is used to group the hotspot snippets into a small number of clusters containing geometrically similar hotspots. 
Given a set of hotspot patterns, a distance metric is defined to calculate the similarity of different hotspot patterns. All patterns are assigned to a cluster where the distance between the cluster center and the pattern is less than the cluster radius. Once the clusters are determined, it is analyzed to produce a description of this cluster, including a representative hotspot snippet and a radius that characterizes the cluster tightness. The representative hotspot in each cluster is then identified and stored in a hotspot library for future hotspot detection. 
The clustering method is further extended in \cite{HOT_DAC2012_Guo} using an improved tangent space based distance metric to achieve better accuracy.

\section{Lithography Hotspot Mitigation} \label{sec:mitigation}

\subsection{Lithography Friendly Placement}

For CMOS feature size significantly smaller than the lithography wavelength (193nm), the printability of a standard cell could be well affected by its neighboring cells.
To minimize the interference between adjacent cells, standard cell design itself and the standard placement methodology need to be co-designed to avoid newly generated lithography hotspots between these cells after placement.
Dummy poly or metal lines may be inserted to create regular neighborhood patterns between neighboring cells.

As the feature size and pitch become even smaller, double or multiple patterning is needed to extend the 193nm lithography. 
A grand challenge for standard cell and placement co-optimization is that there may be coloring conflict between adjacent cells.
In fact, it is still an open question whether the cells shall be pre-colored (i.e., during the standard cell layout stage)
or post-colored (i.e., flat after standard cell placement of of the entire chip).
Liebmann et al. in \cite{PLACE_SPIE2011_Liebmann} proposed some guidelines to enable double patterning friendly standard cell design and placement.
Other placement studies toward double patterning are present \cite{PLACE_ICCAD09_Gupta}\cite{PLACE_SPIE2013_Gao}.
Recently, Yu et al. \cite{PLACE_ICCAD2013_Yu} proposed a systematic framework to seamlessly integrate triple patterning constraints for standard cell and placement stages.

\subsection{Lithography Friendly Routing}
 
Lithography hotspot mitigation can be performed at the post-routing stage, e.g., \cite{ROUTING_DAC05_Mitra}.
In \cite{HOT_SPIE10_Yang}\cite{HOT_SPIE12_Gao}, design rule checker is integrated with the routing engine at the post-routing stage to identify and correct hotspots. First, a set of problematic pattern topologies are transformed into DRC rules. Once placement and routing is done, those pre-built rules are applied to the layout to identify violated pattern region based on a pattern matching based rule checker. These approaches can provide a fast feedback to the router on the hotspot location, and then the router can apply rip-up and reroute to fix the hotspots. 

However, fixing hotspots at this post-routing stage has limited flexibility as only limited rip-up and reroute may be performed. With efficient hotspot predictions, it will be interesting to integrate lithography hotspot detection together with routing. 

\begin{figure}[b]
 \centering
\includegraphics[width=3.3in]{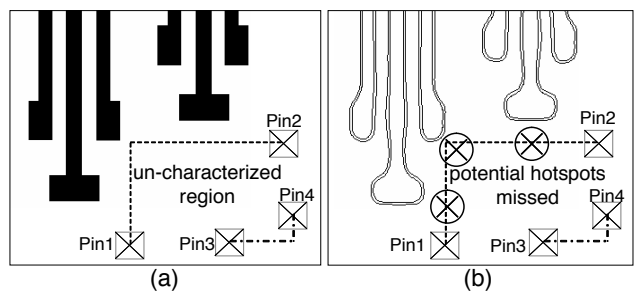}
\vspace{-0.1in}
  \caption{The hotspot detection challenge in the detailed routing stage \cite{ding_aeneid:_2011}.}
  \label{figure:routing_dilemma}
\end{figure}

One challenge of lithography-aware routing is that hotspots are difficult to be detected before a real routing path is obtained. Fig. \ref{figure:routing_dilemma}(a) shows a layout region with metal blockages and unrouted pins Pin1-Pin4. Because some nets are not yet routed, there is an un-characterized region where no hotspots would be identified by general hotspot detection methods. Consequently, potential hotspots may be caused by route Pin1-Pin2 as shown in Fig. \ref{figure:routing_dilemma}(b). Ding et al. \cite{ding_aeneid:_2011} proposed a lithography-friendly detailed routing based on a pre-built hotspot prediction kernel and a routing path prediction kernel. First, the hotspot detection kernel is trained to evaluate the pattern printability based on a set of post-RET data. To overcome the issue of un-characterized regions, the routing path prediction kernel is established using the following steps: (1) explore the possible routing solutions given the available routing resources; (2) perform accurate lithography simulation for the possible layout results; (3) identify preferable routes according to results of hotspots and routing congestion. Because the data that need to be processed for building the routing path prediction kernel is huge, a neural network classifier is constructed to guide the routing engine. The experimental results showed very promising results with this approach.

\section{Conclusion} \label{sec:conclude}
\label{sec:conc}
Lithography hotspots have a great impact on the manufacturing yield. Identifying these problematic layouts during physical design has become a critical problem. Since full chip lithography simulation is computational expensive, pattern matching and machine learning based hotspot detection are very useful for full-chip scale physical verification/screening and layout optimization. In this paper, we discuss some key issues of the lithography hotspot detection problem, including critical feature extraction, pattering matching, and machine learning based methods. We also discuss hotspot mitigation, e.g., at placement and routing stages. It shall be noted that the accuracy and robustness of hotspot detection, and the integration with physical design still have a lot of room for improvement and future research, in particular for multiple patterning and other emerging lithography technologies. 


\section*{Acknowledgment}
This work is supported in part by NSF grants CCF-0644316 and
CCF-1218906, SRC task 2414.001, NSFC grant 61128010, IBM, and Oracle.

\bibliographystyle{IEEEtran}
\bibliography{./Ref/top,./Ref/HSD,./Ref/mitigation}

\end{document}